\documentclass[iopart,superscriptaddress,amsfonts,amsmath,amssymb,showpacs,twocolumn,floatfix,nobalancelastpage]{revtex4-1}
\pdfoutput=1
\usepackage{url}
\usepackage{bm}
\usepackage{graphicx}
\usepackage{amsmath}
\usepackage{amstext}
\usepackage{amssymb}
\usepackage{amsfonts}
\usepackage{amsbsy}
\usepackage{verbatim}
\usepackage{color}
\usepackage[colorlinks=true, urlcolor=blue, linkcolor=blue, citecolor=blue, pdftex]{hyperref}

\begin{document}

\title{Entanglement Hamiltonian of the quantum N\'eel state}

\author{Didier \surname{Poilblanc} }
\affiliation{Laboratoire de Physique Th\'eorique, CNRS, UMR 5152 and Universit\'e de Toulouse, UPS,
F-31062 Toulouse, France}

\date{\today}

\begin{abstract} 
Two-dimensional Projected Entangled Pair States (PEPS) provide a unique framework giving access to detailed entanglement features of correlated (spin or electronic) systems. For a bi-partitioned quantum system, it has been argued that the Entanglement
Spectrum (ES) is in a one-to-one correspondence with the physical edge spectrum on the cut 
and that the structure of the corresponding 
Entanglement Hamiltonian (EH) reflects closely bulk properties (finite correlation length, criticality, topological order, etc...). 
However, entanglement properties of systems with spontaneously broken continuous symmetry are still not fully understood.
The spin-1/2 square lattice Heisenberg antiferromagnet provides a simple example showing spontaneous breaking of SU(2) symmetry down to U(1). The ground state can be viewed as a ``quantum N\'eel state" where the classical (N\'eel) staggered magnetization is reduced by quantum fluctuations.
Here I consider the (critical) Resonating Valence Bond state doped with spinons to describe such a state, that
enables to use the associated PEPS representation (with virtual bond dimension $D=3$) to compute the EH and the   ES for a partition of an (infinite) cylinder. In particular, I find that 
the EH is (almost exactly) a chain of a dilute mixture of heavy  
($\downarrow$ spins) and light ($\uparrow$ spins) hardcore bosons, where light particles
are subject to long-range hoppings. The corresponding ES shows drastic differences with the typical ES obtained previously for ground states with restored SU(2)-symmetry (on finite systems).   

\end{abstract}
\maketitle
\section{Introduction}

It is known from early Quantum Monte Carlo (QMC) simulations that the ground state (GS) of the spin-1/2 Heisenberg antiferromagnet (AFM) on the bipartite square lattice is magnetically ordered~\cite{QMC_young}
and, hence, breaks the hamiltonian SU(2) symmetry.
The GS can be viewed as a ``quantum N\'eel state" (QNS) where the maximum classical value $m_{\rm stag}=1/2$
of the staggered magnetization is reduced by (moderate) quantum fluctuations. More recent QMC 
simulations~\cite{QMC_sandvik} have provided GS energy, staggered magnetization and spin-spin correlations with unprecedented accuracy. In particular,
it has been established that the QNS exhibits power-law decaying spin-spin correlations 
characteristic of a {\it critical state}~\cite{QMC_sandvik2}. 

Recently, a number of new powerful tools based on entanglement measures have emerged.
The entanglement spectrum (ES) and its associated Entanglement Hamiltonian defined
via the reduced density matrix (RDM) of a bi-partioned quantum system (see definitions later)
 provides new insights. In particular, it has been argued that the ES is in a one-to-one correspondence with the physical edge spectrum on the cut
for topological ground states~\cite{li_haldane} and low-dimensional quantum antiferromagnets~\cite{didier}
and that the structure of the corresponding 
Entanglement Hamiltonian (EH) reflects closely the bulk properties (holographic principle)~\cite{PEPS_cirac}. However, new interesting features might  arise in the entanglement properties of systems with spontaneously broken continuous symmetry, such as the QNS for which SU(2) symmetry is broken down to U(1).
First, the entanglement entropy (the entropy associated to the RDM)
have revealed anomalous additive (logarithmic) corrections~\cite{ee,ee2} 
to the area law -- i.e. the linear (asymptotic) scaling of the entropy with the length of the cut. It was proposed afterwards that the origin of such corrections may lie in the existence of Goldstone modes~\cite{metlitski} associated to the 
spontaneously broken continuous symmetry. Note however that, in any finite system (as in most ``exact'' simulations), the SU(2) symmetry is restored by quantum fluctuations and one has a unique GS instead of a degenerate manifold.
In fact, recent state-of-the-art SU(2)-symmetric Density Matrix Renormalization Group (DMRG) studies established an interesting 
correspondence~\cite{TowerStates}, in the (singlet) ground state of two-dimensional 
antiferromagnets in their magnetically-ordered phases, between the SU(2) tower of states and the lower part of the ES below an ``entanglement gap'' (although DMRG does not provide information on the momenta of the ES). 
This suggests strongly that the above-mentioned corrections in the entropy should be associated with the tower of states structure, while the area law arises from ES levels above the entanglement gap~\cite{TowerStates}.
{\it A priori} important differences may occur in the entanglement properties 
of a N\'eel-like wave-function breaking the continuous SU(2) symmetry explicitly i.e. with a finite staggered magnetization.
In particular, one expects the ES (and the EH) of a symmetry-broken QNS 
to differ qualitatively from the ones associated to the GS with restored SU(2) symmetry,
computed on finite systems~\cite{TowerStates,EH_luitz}.  
Computing the entanglement properties of a (variational) state with a finite order parameter is the main goal of this paper. 

\begin{figure}
\includegraphics[width=0.9\columnwidth,clip]{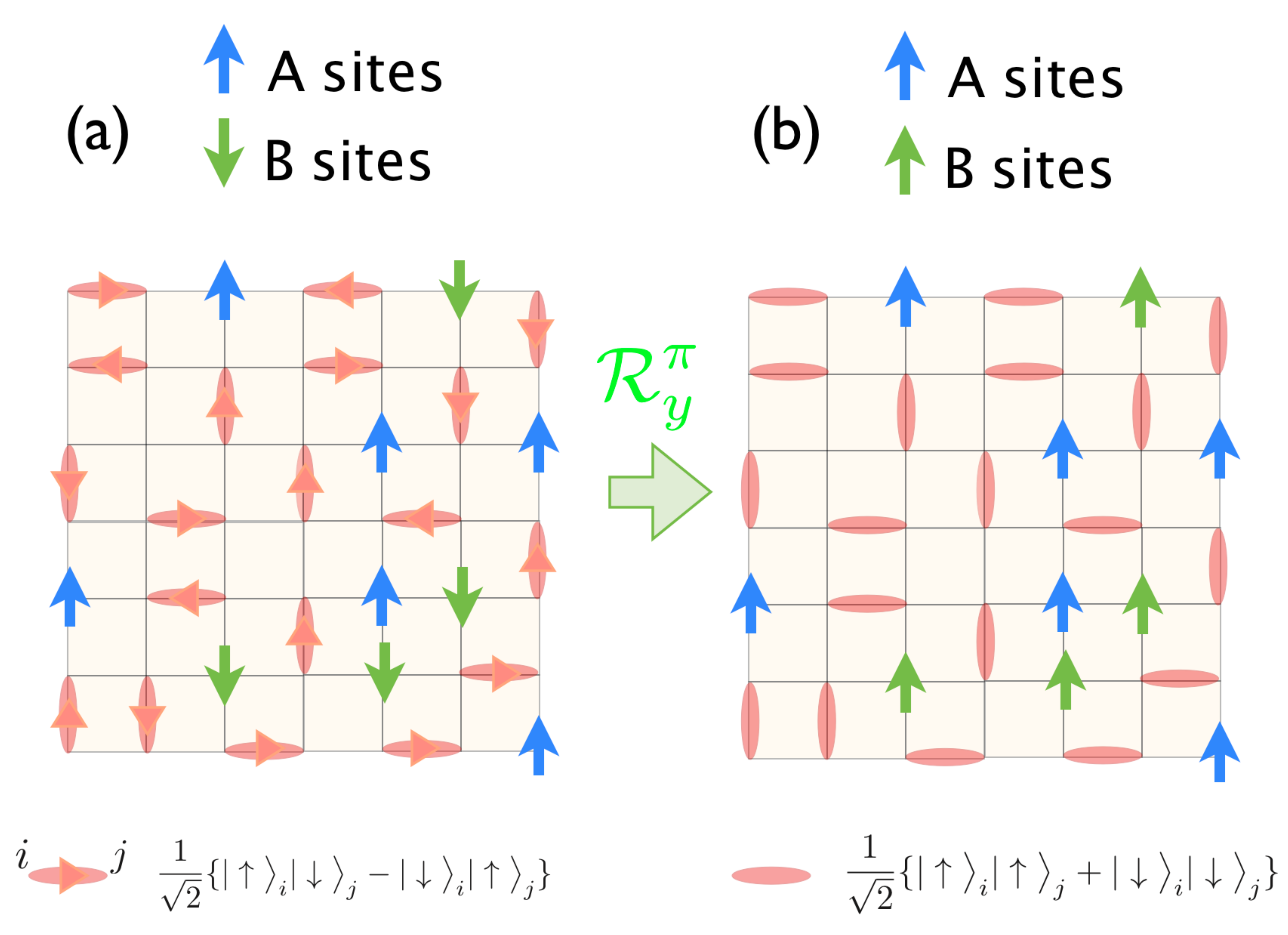}
\caption{(a) The N\'eel state is represented as a spinon-doped RVB state~: Singlets are oriented from the A to the B sublattice
and doped spinons are polarized along $\hat z$ ($-\hat z$) on the A (B) sites. Implicitly, a sum over all singlet/spinon configurations is assumed, the average spinon density being controlled by a fugacity.
(b) Under a $\pi$-rotation around  
$\hat y$ on all the B-sites, all spinons become oriented along $\hat z$ and singlets transform into \hbox{$\frac{1}{\sqrt{2}}
\{|\uparrow\uparrow\big> + |\downarrow\downarrow\big> \}$} on every NN bonds.}
\label{Fig:neel}
\end{figure}

The formalism of Projected Entangled Pair States (PEPS)~\cite{verstraetewolf06,PEPS_cirac} 
enables i) to easily construct symmetry broken variational states and ii) to compute the corresponding EH. 
Note that, for a given variational state and system size (one uses infinite cylinders with a finite perimeter), the calculation of the EH is {\it fundamentally exact} and provides a complete analytic expansion in terms of N-body interactions whose amplitudes are numerically computed.
Here I therefore make use of a simple PEPS ansatz of the QNS in order to calculate its EH associated to a bi-partition of an infinite cylinder. The variational wave function used here is in fact the simplest PEPS (i.e. with 
the smallest bond dimension $D=3$) one can construct to capture the physics of the symmetry-broken N\'eel state. 
Ans\"atze with a larger bond dimension will not allow to consider a cylinder with a large enough perimeter. 
Note that the PEPS formalism provides also the momentum-resolved ES. This is to be contrasted to DMRG
that also gives easily the ES but without the corresponding momenta of the Schmidt states. 
Also an analytic form of the EH cannot be obtained in DMRG. 

As shown recently using PEPS, a EH with local 
interactions is expected in a gapped 
bulk phase (with
short-range entanglement), whereas 
a diverging interaction length of the EH is the hallmark of critical behavior
in the bulk~\cite{PEPS_cirac}.
One therefore expects to see fingerprints of the critical behavior of the QNS in its
Entanglement Hamiltonian.

\section{Doped-RVB ansatz for the N\'eel state}

I start with the square lattice Resonating Valence Bond (RVB) wavefunction defined as an equal-weight superposition of nearest-neighbor (NN)  hardcore singlet coverings~\cite{RVB_anderson1,RVB_anderson2}.
The sign structure of the wave function is fixed by imposing that the singlets
\hbox{$|\uparrow\downarrow\rangle-|\downarrow\uparrow\rangle$} are all oriented from one A sublattice to the other B sublattice. Such a wave function is a global spin singlet -- i.e. a SU(2)-invariant state -- with 
algebraic (i.e. critical) dimer correlations (and short-range spin correlations)~\cite{RVB_critical1,RVB_critical2}.
To construct a simple ansatz for the QNS, let us now assume 
that  one breaks SU(2) symmetry down to U(1) by doping the NN
RVB state with on-site spinons (i.e. spin-1/2 excitations) with {\it opposite} orientations on the two sublattices.
For simplicity, I choose hereafter the staggered magnetization pointing along the $\hat z$-axis. 
Such a simple ansatz is schematically shown in Fig.~\ref{Fig:neel}(a).  
The average density of spinons -- identical on the two sublattices -- directly gives the staggered magnetization $m_{\rm stag}$ ($\times 2$) and, as one will see later on, can be controlled by a fugacity $\gamma$. 

Before going further, it is convenient to rewrite the N\'eel-RVB state in a translationally invariant form.
Indeed, under a (spin) $\pi$-rotation around  
$\hat y$ on the B-sites, B-spinons transform as \hbox{$|\downarrow\big>\rightarrow|\uparrow\big>$} and 
\hbox{$|\uparrow\big>\rightarrow-|\downarrow\big>$}. Under such a (unitary) transformation, the new 
N\'eel-RVB state acquires the same (average)
polarization on the A and B sublattices as shown in Fig.~\ref{Fig:neel}(b).
The original NN singlets are also transformed into \hbox{$\frac{1}{\sqrt{2}}
\{|\uparrow\uparrow\big> + |\downarrow\downarrow\big> \}$} dimers which are now symmetric w.r.t. the bond centers.

\section{PEPS construction and energetics}

\begin{figure}
\begin{center}
 \includegraphics[width=0.9\columnwidth,clip]{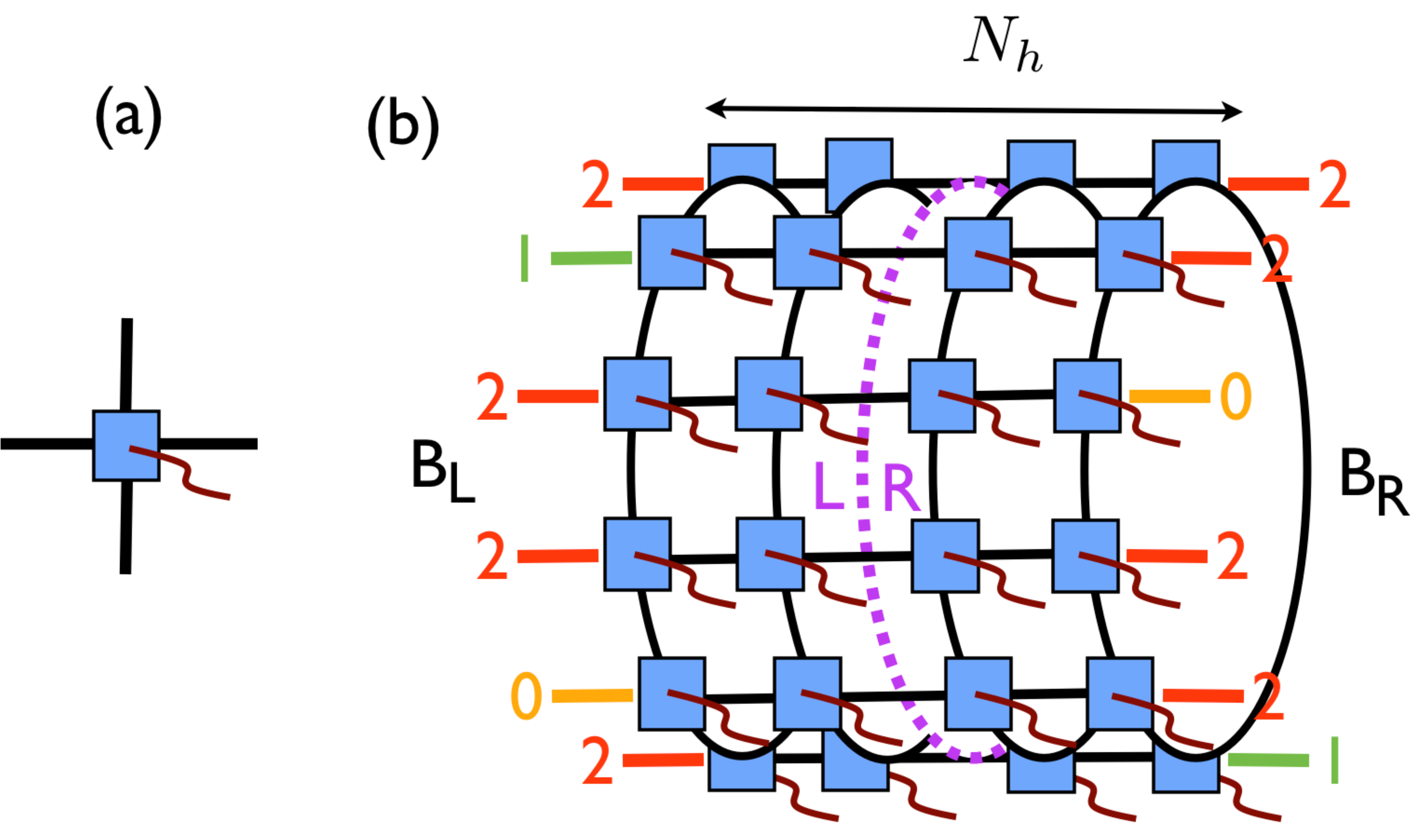}
 \end{center}
\caption{(Color online)
(a) Local (rank-5) PEPS tensor. (b) Tensors are placed on a square lattice wrapped on a cylinder of perimeter $N_v$
and (quasi-) infinite length $N_h\gg N_v$. $B_L$ and $B_R$ boundary conditions 
are realized by fixing the virtual variables going out of the cylinder ends. A bipartition of the cylinder
generates two L and R edges along the cut. }
\label{Fig:peps}
\end{figure}

Such a state can in fact be
represented by a PEPS $|\Psi_{\rm PEPS}\big>$ with bond dimension $D=3$, where
each lattice site is replaced by a rank-5 tensor ${\cal A}^{s}_{\alpha,\alpha';\beta,\beta'}$
labeled by one physical index, $s=0$ or $1$,
and by four virtual bond indices (varying from 0 to 2) along the horizontal ($\alpha,\alpha'$) and vertical ($\beta,\beta'$) directions, as shown in Fig.~\ref{Fig:peps}(a).
Physically, the absence of singlet on a bond is encoded by the virtual index being "2" on that bond. 
I define~: 
\begin{equation}
{\cal A}={\cal R+\gamma S}  \, ,
\end{equation}
where $\cal R$ is the original RVB tensor~\cite{RVB_norbert,RVB_didier}, $\cal S$ is  
a polarized spinon tensor and $\gamma\in \mathbb{R}$ is a fugacity controlling the average spinon density. 
To enforce the hardcore dimer constraint, one takes
${\cal R}^{s}_{\alpha,\alpha';\beta,\beta'}=1$ whenever three virtual indices equal 2
and the fourth one equals $s$, and ${\cal R}^{s}_{\alpha,\alpha';\beta,\beta'}=0$ otherwise. 
The spinon tensor has only one non-zero element, ${\cal S}^{1}_{2,2;2,2}=1$.
The wave function amplitudes are then
obtained by contracting all virtual indices (except the ones at the boundary of 
the system). Note that the above PEPS ansatz for the N\'eel state bares similarities with the one 
used to describe the honeycomb RVB spin liquid under an applied magnetic field~\cite{RVB_magnet}.
However, a crutial difference is that this new ansatz is, by construction, fully U(1)-invariant in contrast to the 
spinon-doped RVB state of Ref.~\cite{RVB_magnet}.

Following a usual procedure, I now place the square lattice of tensors on infinite cylinders with $N_v$ 
sites in the periodic (vertical) direction as shown in Fig.~\ref{Fig:peps}(b) and use standard techniques (involving 
exact tensor contractions and iterations of the transfer operator) to compute relevant observables.
In the PEPS formulation the boundary conditions $B_L$ and $B_R$ can be simply set by fixing the 
virtual states on the bonds ``sticking out" at each cylinder end.
E.g. open boundary conditions are obtained by setting 
the boundary virtual indices to ``2". Generalized boundary conditions can be realized as in Fig.~\ref{Fig:peps}(b) by setting some of the virtual indices on the ends to ``0" or ``1".

\begin{figure}
\begin{center}
 \includegraphics[width=0.9\columnwidth,clip]{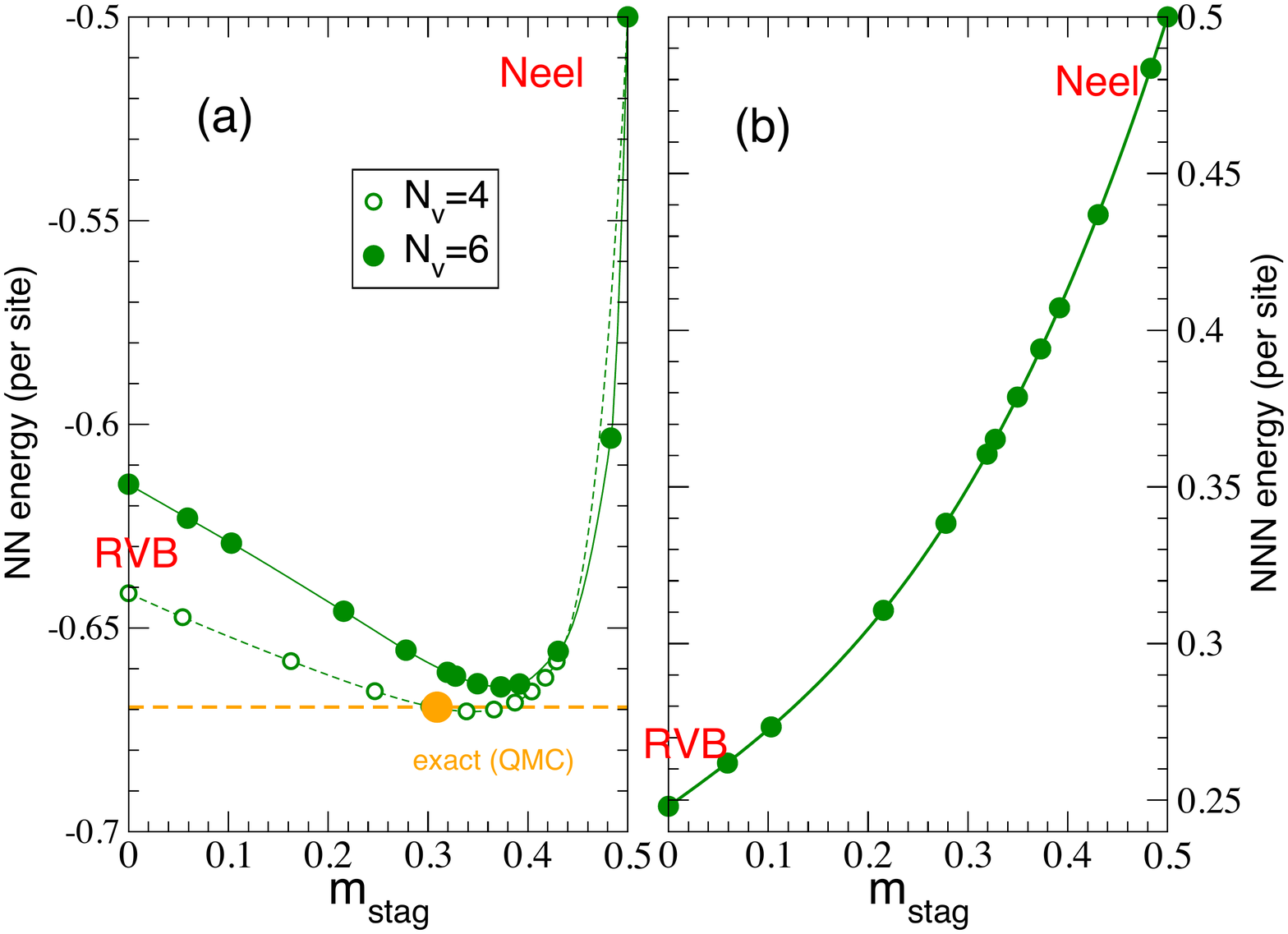}
 \end{center}
\caption{(Color online)
NN (a) and next-NN (b) correlators $2\big<{\bf S}_i\cdot{\bf S}_j\big>$
-- corresponding to the energies per site in units of the coupling constants -- 
plotted as a function of $m_{\rm stag}$. Computations are done on infinite cylinders
of perimeter $N_v=4$ and $N_v=6$. }
\label{Fig:energies}
\end{figure}

I have computed the (staggered) magnetization $m_{\rm stag}$ and the expectation values 
of the \hbox{spin-1/2} Heisenberg exchange interactions ${\bf S}_i\cdot{\bf S}_j$ between NN and next-NN sites,
varying $\gamma$ from zero to large values (to approach the classical N\'eel state).
The data (normalized as the energy per site of the corresponding Heisenberg model) 
are displayed as a function of $m_{\rm stag}$ in Fig.~\ref{Fig:energies}(a,b).
The NN energy shows a broad minimum around $m_{\rm stag}\sim 0.35$, a value a bit larger than the 
QMC extrapolation $\sim 0.307$~\cite{QMC_sandvik} 
for the pure NN quantum AFM. 
However, (i) the variational energy curve is rather flat around the minimum and (ii) the minimum energy 
is only within $\sim 1.5\%$ of the QMC estimate, a remarkable result considering the 
simplicity of the one-dimensional family of $D=3$ PEPS. Note also that the minimum
energy agrees very well with optimized $D=3$ iPEPS~\cite{bauer} and finite PEPS up to $D=6$~\cite{michael}. 

For completeness, I also show the next-NN energy in Fig.~\ref{Fig:energies}(b). In fact, the pure (critical) RVB 
state provides the lowest next-NN exchange energy, suggesting the existence of a transition, upon increasing the next-NN coupling, from the N\'eel state to a {\it gapless} spin liquid ~\cite{wang,sheng}. Note that a direct transition from the N\'eel state to a Valence Bond Crystal -- with no intermediate gapless spin liquid phase -- is also a realistic scenario~\cite{vbc,QMC_sandvik2}.

\section{Entanglement Hamiltonian on infinite cylinders} 
\label{Sec:boundary}

\subsection{Bipartition and reduced density matrix} 

To define an Entanglement Hamiltonian associated to the family of N\'eel-RVB wavefunctions,
I partition the $N_v\times N_h$ cylinder into two half-cylinders
of lengths $N_h/2$,  as depicted in Fig.~\ref{Fig:peps}(b). 
Partitioning the cylinder into two half-cylinders
reveals two edges L and R along the cut. Ultimately, I
aim to take the limit of infinite N\'eel-RVB cylinders, i.e. $N_h\rightarrow\infty$ as before. 

The reduced density matrix of the left half-cylinder obtained by tracing over the degrees of freedom of the right half-cylinder,
$\rho_L={\rm Tr}_R\{
|\Psi_{\rm PEPS}\big>\big<\Psi_{\rm PEPS}|\}$, can be simply mapped, via a spectrum conserving 
isometry $U$, onto an operator $\sigma_b^2$ acting only on the $D^{\otimes N_v}$ edge (virtual) degrees of freedom, i.e. $\rho_L=U^\dagger \sigma_b^2 \,U$~\cite{PEPS_cirac}.  
The {\it Entanglement (or boundary) Hamiltonian} $H_b$ introduced above is defined as $\sigma_b^2=\exp{(-H_b)}$. 
As $\sigma_b^2$, $H_b$
is one-dimensional and its spectrum -- the {\it entanglement spectrum} (ES) -- is the same as the one of $-\ln{\rho_A}$.
Note that the left and the right half-cylinders give identical EH. 
For further details on the derivation and the procedure, the reader is kindly asked to refer 
to Ref.~\onlinecite{PEPS_cirac}.

For a topological state, such as the $\gamma=0$ RVB state, the Entanglement Hamiltonian 
depends on the choice of the
$B_L$ and $B_R$ cylinder boundaries that define ``topological sectors"~\cite{RVB_didier,Topo_norbert}.
Adding any staggered magnetization $m_{\rm stag}$ in the PEPS
immediately breaks the gauge symmetry of the tensors
which is responsible for the disconnected topological sectors, as also happens in the case of field-induced
magnetized RVB states~\cite{RVB_magnet}. 
Therefore, all topological sectors are mixed and $H_b$ become independent of
 the boundary conditions $B_L$ and $B_R$ provided $N_h\rightarrow\infty$.
 Note also that $H_b$ inherits the U(1) symmetry (associated to rotations around the direction
 of $m_{\rm stag}$) of the N\'eel state. 

\subsection{Expansion in terms of N-body operators}

To have a better insight of the Entanglement Hamiltonian, I expand it in terms of a basis of $N$-body operators, $N=0,1,2,\cdots$~\cite{PEPS_cirac,RVB_didier}.
For this purpose, I use a local basis of $D^2=9$ (normalized)
${\hat x}_\nu$ operators, $\nu=0,\cdots,8$ which act 
on the local (i.e. at some site $i$) configurations $\{ |0\big>, |1\big>, |2\big> \}$, where 
$|2\big>$ is the vacuum or ``hole" state and $|0\big>$ and $|1\big>$ can be viewed as 
spin down and spin up particles, respectively. More precisely,
${\hat x}_0=\mathbb{I}^{\otimes 3}$, ${\hat x}_1=\sqrt{\frac{3}{2}}(|0\big>\big<0| -|1\big>\big<1|)$ 
and ${\hat x}_2=\frac{1}{\sqrt{2}}(|0\big>\big<0|+|1\big>\big<1|-2|2\big>\big<2|)$,
for the diagonal matrices, complemented by  
$\hat x_3=\hat x_4^\dagger=\sqrt{3}|0\big>\big<1|$
acting as (effective) spin-1/2 lowering/raising operators,
and $\hat x_5=\hat x_7^\dagger=\sqrt{3} |2\big>\big<0|$ and 
$\hat x_6=\hat x_8^\dagger=\sqrt{3} |2\big>\big<1|$ acting as particle hoppings. 
In this basis $H_b$ reads~\cite{RVB_didier},
\begin{eqnarray}
H_b &=& c_0 N_v+\sum_{\nu,i} c_{\nu} {\hat x}_\nu^i 
+ \sum_{\nu,\mu,r,i} d_{\nu\mu}(r) \, {\hat x}_\nu^i {\hat x}_\mu^{i+r} 
\nonumber \\
&+& \sum_{\lambda,\mu,\nu,r,r',i} e_{\lambda\mu\nu}(r,r') \, {\hat x}_\lambda^i {\hat x}_\mu^{i+r} {\hat x}_\nu^{i+r'} 
+ \cdots \, ,\label{Eq:Hb0}
\end{eqnarray}
where site superscript indices have been added and only the first one-body, two-body and three-body terms are shown.

\begin{figure}\begin{center}
 \includegraphics[width=0.9\columnwidth]{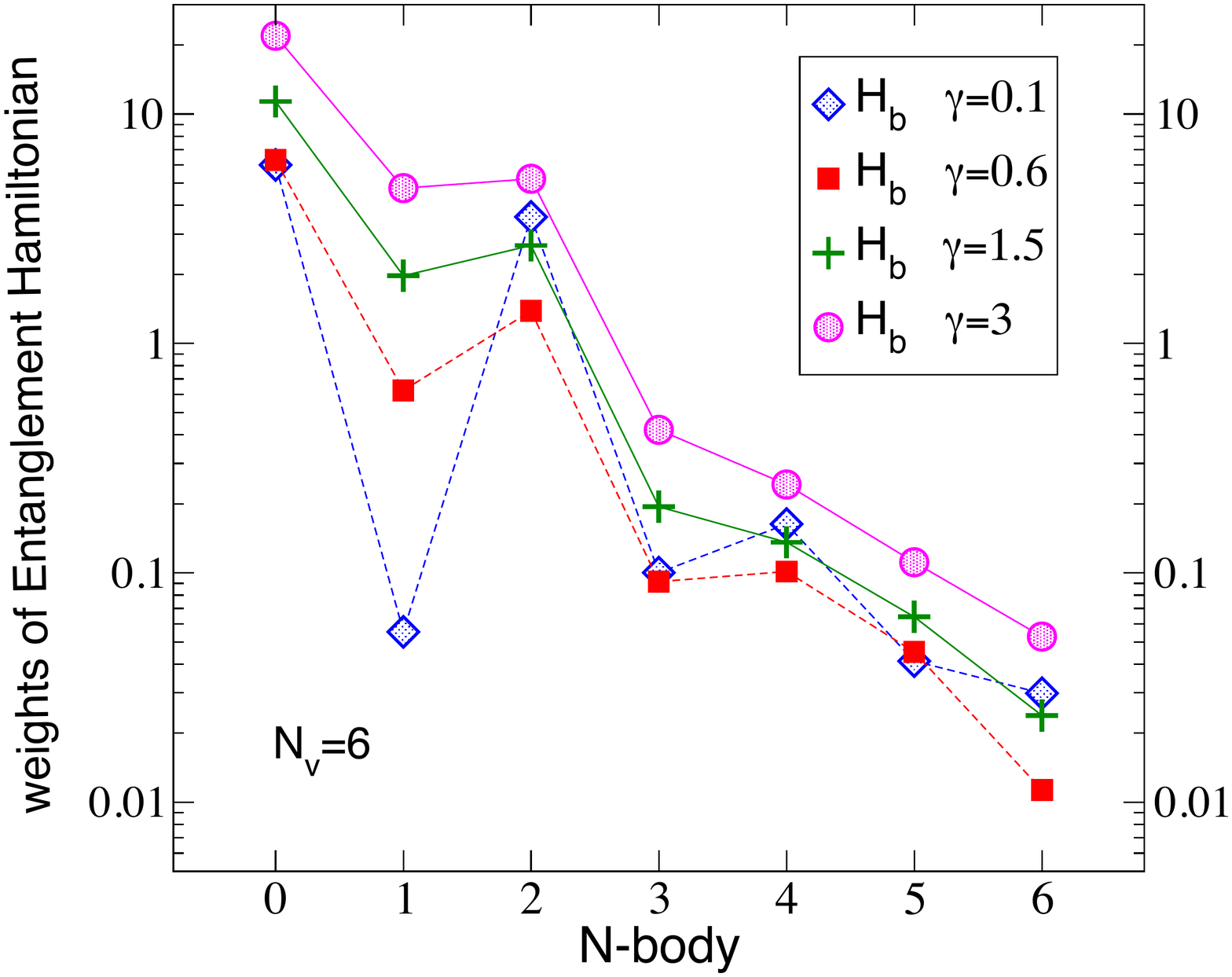}
 \end{center}
\caption{(Color online)
Weights of the Entanglement Hamiltonian $H_b$
expended in terms of N-body operators. Data of several N\'eel-RVB wavefunctions
(whose $\gamma$ values are mentioned on the plot) are shown. Calculations are done
on an infinite cylinder with perimeter $N_v=6$. As seen e.g. in Ref.~\protect\cite{RVB_didier},
finite size effects for such integrated quantities are typically quite small.
}
\label{Fig:weights}
\end{figure}

The total weights corresponding to each order of the expansion of $H_{b}$ in terms of N-body 
operators are shown in
Fig.~\ref{Fig:weights} as a function of the order $N$ using a semi-logarithmic scale.  
The data reveal clearly a fast decay of the weight
with the order $N$. This decay is compatible with an exponential law although more decades
in the variation of the weights (i.e. larger $N_v$) would be needed to draw a definite conclusion. 
In any case, $H_{b}$ is dominated by two-body contributions
in addition to the normalization constant and subleading one-body terms. 
The quantum N\'eel state is believed to be critical with 
power-law decay of spin-spin correlations~\cite{QMC_sandvik2}. Therefore, according to Ref.~\cite{PEPS_cirac}, 
one expects $H_b$ to 
be long-ranged to some degree. So, one still needs to refine the analysis
and investigate further the r-dependence of the leading two-body contributions.
In the next Subsection, I show that $H_b$ indeed possesses long-range two-body terms that I characterize.  

\subsection{Entanglement Hamiltonian: an effective one-dimensional t--J model}

It is known that the EH of the $\gamma=0$ RVB PEPS belongs to the $1/2\oplus 0$ representation
of SU(2) and its Hilbert space is the same as the one of a one-dimensional bosonic 
\hbox{t--J} model~\cite{RVB_didier}, interpreting $|0\big>$ and $|1\big>$ states ($|2\big>$ states) as $\downarrow$ and 
$\uparrow$ spins (holes).
In the presence of a finite (staggered) magnetization in the bulk, the SU(2) symmetry is broken but $H_b$ keeps 
the unbroken $U(1)$ symmetry corresponding to spin rotations around the direction of the staggered magnetization.

\begin{figure}
\includegraphics[width=0.9\columnwidth]{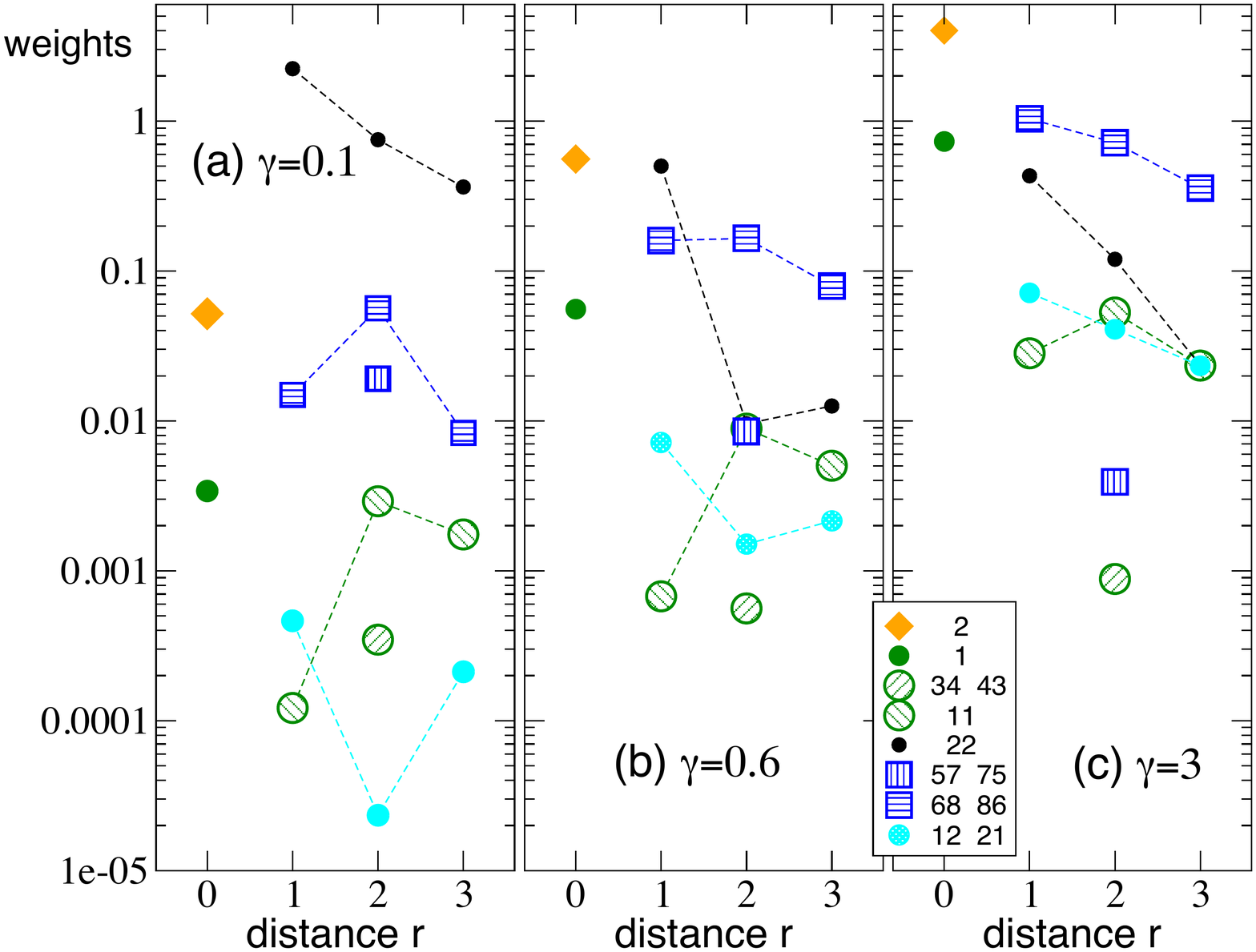}
\caption{Largest weights $|c_\nu|$ and $|d_{\nu\mu}(r)|^2$ of the one-body (i.e. $r=0$) and two-body operators
in the expansion of $H_b$ of the N\'eel-RVB PEPS as a function of distance $r$, for increasing $\gamma$ values
(corresponding to staggered magnetizations $m_{\rm stag}\sim 0.059, 0.327$ and $0.483$, respectively).}
\label{Fig:bham}
\end{figure}

The (largest) non-zero real coefficients in (\ref{Eq:Hb0}) computed 
on an infinitely-long cylinder of perimeter $N_v=6$ are shown in Fig.~\ref{Fig:bham}(a-c) for small (a), intermediate (b)
and large (c) (staggered) magnetizations.
At large and intermediate values of $m_{\rm stag}$, one finds a dominant one-body (diagonal) term which can be interpreted as 
a chemical potential term (up to a multiplicative factor)~:
\begin{equation}
{\cal H}_2=c_2\sum_i {\hat x}_2^i = \frac{3}{\sqrt{2}}c_2\sum_{i}  (n_i-2/3)  \, ,
\label{Eq:mu}
\end{equation}
where $n_i$ counts the number of particles (i.e. ``0" and ``1" states) on site $i$.
The subleading one-body operator takes the form of a Zeeman coupling~:
\begin{eqnarray}
{\cal H}_{1}&=&\sqrt{6}c_1\sum_{i} S_i^z\, ,
\label{Eq:nn}
\end{eqnarray}
where $S_i^z$ is an effective spin-1/2 component (along $\hat z$) and $c_1\simeq c_2/10$. 

The leading 2-body contributions are
hopping terms {\it at all distances} for the majority ``spins" ($|1\big>$ states)~:
\begin{eqnarray}
{\cal H}_{68}(r)&=&d_{68}(r)\sum_{i} ({\hat x}_8^i{\hat x}_6^{i+r} + {\hat x}_6^i{\hat x}_8^{i+r})\nonumber \\
&=&3d_{68}(r)\sum_{i} (b_{i+r,1}^\dagger b_{i,1} + b_{i,1}^\dagger b_{i+r,1}) \, , 
\label{Eq:hopping}
\end{eqnarray}
where $b_{i,s}^\dagger$ ($b_{i,s}$) are the canonical bosonic creation (annihilation) operators of the virtual $s=0,1$ states.
Note that the minority spins ($|0\big>$ states) only hop at {\it even} distances (weights at odd distances are 
negligable) with much weaker amplitudes,
\hbox{$d_{57}=d_{75}\ll d_{68}=d_{86}$}. 
The next subleasing corrections are diagonal 2-body density-density interactions 
\begin{eqnarray}
{\cal H}_{22}(r)&=&d_{22}(r)\sum_{i} {\hat x}_2^i{\hat x}_2^{i+r}\nonumber \\
&=&\frac{9}{2}d_{22}(r)\sum_{i} (n_i-2/3) (n_{i+r}-2/3)\, ,
\label{Eq:nn}
\end{eqnarray}
which become dominant when $\gamma, m_{\rm stag}\rightarrow 0$.

Other generic operators allowed by the $U(1)$ symmetry, like the anisotropic XXZ chain 
($d_{11}\ne d_{34}=d_{43}$) or mixed operators of the form ${\cal H}_{12}\propto
\sum_{i} S_i^z (n_{i\pm r}-2/3)$
are also present but their amplitudes turn out to be quite small.
Interestingly, $H_b$ (approximately) conserves the hole ``2-charge" and, hence, does not contain pair-field
operators with sizable amplitudes,
in contrast to previous studies of $D=3$ PEPS~\cite{RVB_didier,RVB_magnet}.
If one restricts to the dominant contributions (\ref{Eq:mu}) and
(\ref{Eq:hopping}), $H_b$ is exactly a chain of a dilute mixture of heavy  
($\downarrow$ spins
or $|0\big>$ states) and light ($\uparrow$ spins or $|1\big>$ states) hardcore bosons, where light particles
are subject to long-range hopping.

\section{\label{sec:ES}
Entanglement spectrum} 

\begin{figure}
\vskip 0.3truecm
\includegraphics[width=0.9\columnwidth,clip]{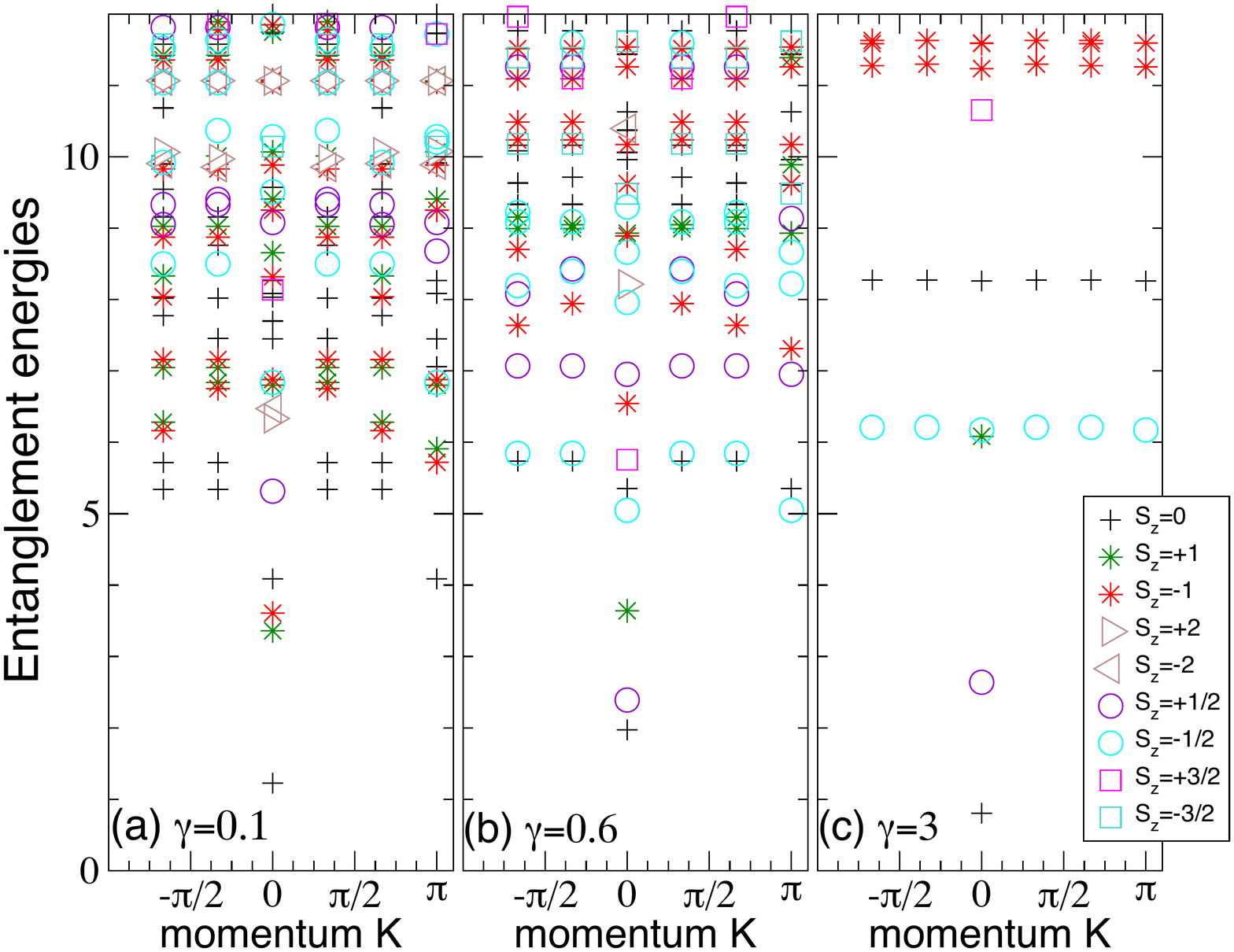}
\caption{Entanglement spectrum of a bipartitioned $N_v=6$ RVB-N\'eel cylinder as a function of the 
momentum along the cut, for different values of the spinon fugacity $\gamma=0.1, 0.6$ and $3$ corresponding to
$m_{\rm stag}\sim 0.059, 0.327$ and $0.483$, respectively. Different symbols are used for different $S_z$ sectors of the edge.  }
\label{Fig:es}
\end{figure}

It is also of high interest to examine the ES in the QNS and compare it to ES obtained 
for GS where SU(2)-symmetry is restored on finite size systems~\cite{TowerStates}.
By definition the ES is the spectrum of $-\ln{\rho_L}$. Since $\rho_L$ and $\sigma_b^2=\exp{(-H_b)}$ are related by an isometry, it is also the spectrum of the Entanglement Hamiltonian $H_b$. ES are shown in Fig.~\ref{Fig:es}(a-c)
for 3 values of the fugacity $\gamma$, as a function of the momentum along the cut. 
Since $\sigma_b^2$ conserves the total $S_z$ of the chain ($U(1)$ symmetry), it can be block-diagonalized 
using this quantum number
and the eigenvalues of $-\ln{\sigma_b^2}$ are displayed in each $S_z$ sector separately. 
It can be seen from Fig.~\ref{Fig:es}(a) that the $\gamma=0$ SU(2) spin multiplets are split by a small spinon 
density. For increasing $\gamma$ (i.e. staggered magnetization), the splittings of the Kramer's multiplets 
increase (see Fig.~\ref{Fig:es}(b,c)) due to the relative increase of the amplitudes of the  
SU(2)-symmetry breaking terms like (\ref{Eq:hopping}) in the EH. In the limit of large $\gamma$ where the 
classical N\'eel state is approached, one finds separated bands of energy levels. It may be that the ES is gapped for 
all $\gamma$ but finite size effects remain too large to reach a definite conclusion. In any case, the ES of 
Fig.~\ref{Fig:es} are to be contrasted to the ES obtained in DMRG for GS with restored SU(2)-symmetry (due to the use of  finite size systems). 
Obviously the two types of ES are very different with a SU(2) tower of states structure at low energy for the ES of the singlet GS~\cite{TowerStates} and a U(1) symmetric ES in the (variational) symmetry-broken N\'eel state. 

\section{Summary and discussion}

In this paper, I have investigated entanglement properties of a simple one-dimensional family of PEPS
designed to describe qualitatively the GS of the square lattice
AFM. These ans\"atze exhibit a finite staggered magnetization i.e. they break explicitly the SU(2) symmetry 
down to U(1) and can be studied on infinite cylinders with a finite perimeter. 
The goal of this study is therefore to examine the effects of such a finite order parameter on various entanglement 
properties and compare them to (QMC or SU(2)-symmetric DMRG) studies where symmetry 
is restored in a finite system. Thanks to the PEPS structure, the
Entanglement Hamiltonian associated to a bipartition of the cylinder can be derived exactly 
(for a fixed perimeter). It is found that the EH inherits the U(1) symmetry of the N\'eel state and possesses a very simple structure~: (i) its Hilbert space is the same as the one of a one-dimensional bosonic 
\hbox{t--J} model, interpreting the 3 virtual states on the edge as a $\uparrow$ spin, 
a $\downarrow$ spin and a hole, (ii) when expended using a local basis of operators, it shows dominant two-body interactions and (iii) higher-order operators
(three-body terms and beyond) represent less than 10$\%$ of its total weight.
Examining in details the form of the two-body interactions, I find that the dominant ones are long-range 
hoppings of the majority (let say $\uparrow$) spins. It is however not possible to distinguish a power-law versus an exponential decay of these hopping terms. In any case,
the associated Entanglement Spectrum is found to be qualitatively very different from the ones obtained in
GS with restored SU(2)-symmetry~\cite{TowerStates} (no tower of states structure is found, as suspected). 
Whether the entropy exhibits additive logarithmic corrections as in
Refs.~\cite{ee,ee2} is difficult to answer. The absence of the tower of states in the ES suggests a negative answer.  
However, an hypothetical power-law decay of the hopping terms in the EH (instead of exponential) might lead to some 
additive corrections to the entropy. 
It would be interesting to complement our calculation of the ES using ``conceptually exact" numerical methods (such as
QMC or SU(2)-symmetric DMRG) on large but finite systems, adding a small external staggered field (to produce a finite order parameter), taking the limit of infinite system size first. 

I acknowledge fundings by the ``Agence Nationale de
la Recherche" under grant \hbox{No.~ANR~2010~BLANC~0406-0} and support from the
CALMIP supercomputer center (Toulouse).  I thank Claire for her patience and I am 
indebted to Fabien Alet, Ignacio Cirac, Nicolas Laflorencie, Roger Melko, Anders Sandvik, 
Norbert Schuch and Frank Verstraete for numerous discussions and insightful
comments.


\end{document}